\documentclass[aps,pre,amsmath,amssymb,twocolumn]{revtex4}      
\usepackage{graphicx}
\begin{document}
%\twocolumn[ \hsize\textwidth\columnwidth\hsize\csname
%@twocolumnfalse\endcsname

\title{General features of the energy landscape in Lennard-Jones like   
model liquids}

\author{
         L.~Angelani$^{1,2}$,
         G.~Ruocco$^{1}$,
         M.~Sampoli$^{3,4}$, and
         F.~Sciortino$^{1,2}$
         }
%\address{
\affiliation{
$^1$Dipartimento di Fisica and INFM, Universit\`a di Roma {\em La  
Sapienza}, P. A. Moro 2, I-00185 Roma, Italy
}
\affiliation{
$^2$SMC - INFM Universit\`a di Roma {\em La Sapienza}, P. A. Moro 2,  
I-00185 Roma, Italy
}
\affiliation{
$^3$Dipartimento di Energetica and INFM - Universit\`a di Firenze, Via  
Santa Marta 3, I-50139 Firenze, Italy
}
\affiliation{
$^4$LENS,Via Nello Carrara 1, I-50019 Sesto-Fiorentino, Italy
}
\date{\today}
%\maketitle
\begin{abstract}
Features of the energy landscape sampled by supercooled
liquids are numerically analyzed for several Lennard-Jones like model  
systems. The properties of quasisaddles (minima of the square gradient  
of potential energy $W\!=\!|\nabla V|^2$), are
shown to have a direct relationship with the dynamical behavior,  
confirming that the quasisaddle order extrapolates to zero at the  
mode-coupling temperature $T_{MCT}$. The same result is obtained
either analyzing all the minima of $W$ or the saddles (absolute minima  
of $W$), supporting the conjectured
similarity between quasisaddles and saddles, as far as the
temperature dependence of the properties influencing the slow dynamics  
is concerned. We find evidence of  universality in the shape of the  
landscape: plots for different systems superimpose into master curves,  
once energies and temperatures are scaled by $T_{MCT}$. This allows to establish 
a quantitative relationship between $T_{MCT}$ and potential energy  
barriers for LJ-like systems, and suggests a possible generalization to 
different model liquids.
\end{abstract}
%\pacs{61.20.Ja, 64.70.Pf, 34.20.Mq} %]
\maketitle

\section{Introduction}
The investigation of the topological and metric properties of potential  
energy surface (PES), often referred to as ``energy
landscape'', is a useful and powerful tool for studying slow dynamics  
in condensed matter, especially in those cases where the lack of order  
(as for example in supercooled liquids) inhibits the use of the  
analytical tools pertaining to the crystalline state
\cite{deb_nature,land_angell,land_sastry,land_buc,land_sastry2,land_keyes}.
The PES approach has been successfully applied to the study of many  
different interacting systems (glasses, proteins, sheared materials,  
and so on). The PES approach started with the introduction of the  
fruitful concept of {\it inherent structures} \cite{still1}. 
In the last  years, several steps toward a more detailed description of the  
statistical properties of the PES have been performed, most of them  
pointing toward a better understanding of  the relationship
between the landscape properties and the emergent dynamical behavior of  
the analyzed systems.

Among others, two landscape-based approaches have proven to be  
particularly stimulating. The first one concerns the detailed analysis of  
the {\it inherent structures} (i.e. the configurations at the minima of  
potential energy) visited by the system at different temperatures. This  
method has allowed to clarify many
interesting phenomena, as, for example, the thermodynamic picture of  
the supercooled liquid regime based on the configurational entropy  
\cite{fs_entropy}, the relationship between fragility and properties of  
inherent structures \cite{land_sastry2}, the analysis of diffusion  
processes in terms of visited inherent structures  
\cite{land_keyes,fabr,la_98}, or the interpretation of
the effective fluctuation-dissipation temperature in the
out-of-equilibrium regime in terms of inherent structures visited  
during aging \cite{fs_aging}, only to cite a few. The second approach  
is based on the analysis of the eigenvalues (normal modes) of the  
Hessian at the instantaneous configurations during
the dynamic evolution of the system, from here the name {\it  
instantaneous normal mode approach} (for an introduction and an  
extended application of this method see the works of Keyes and  
coworkers \cite{keyes_inm,keyes_vari}). This approach allowed to
relate the emergent diffusive processes to the features of the  
landscape, opening the way to the interpretation of diffusion in terms  
of accessible paths in the multidimensional energy surface. 
Promising steps was obtained {\it i)}
using simultaneously both the instantaneous normal  
mode approach and the inherent structure one, in order to identify the  
relevant slow diffusive directions \cite{donati,lanave},
and {\it ii)} by analyzing the reaction paths in order to eliminate the non-diffusive 
unstable modes \cite{bembe}.

Recently, a further approach has been introduced
\cite{noi_sad,cav_sad} and applied to the study of supercooled  
liquids\cite{sad_1,doye,sad_3,sad_4,sad_5,grig,sadBLJ}. This approach  
is based on the analysis of the {\it saddles} of the
potential energy surface and has provided new insight in the analysis  
of the dynamic crossover taking place on lowering the temperature in  
supercooled liquids. Indeed it allows to characterize the dynamic  
transition temperature $T_{MCT}$
(mode-coupling temperature \cite{mct}) as the temperature where  
the order (fractional number of negative eigenvalues of the  
Hessian matrix) of the saddles vanishes. 
This finding suggested the following  
scenario for the dynamics: above $T_{MCT}$ the
representative point in the configuration space lies close to the  
saddles and the relevant dynamic process is the diffusion among  
multidimensional saddle points, i.e. the diffusion takes place along  
paths at almost constant potential energy, and the limiting factors to  
particles diffusion are ``entropic'', rather than ``energetic'',  
barriers. On the contrary, below $T_{MCT}$ the
minimum-to-minimum diffusion processes dominate and the ``true''  
barrier jump controls the diffusive dynamics. A clear  
landscape-based interpretation of the dynamic behavior of the system is  
then provided. It is important to mention here that
the term ``saddles'' is not mathematically correct, as the way the  
saddles have been defined in Ref.s~\cite{noi_sad,cav_sad} is based
on the partition of the configuration space in basins of
attraction of the minima of the ``pseudopotential''\cite{still_W}
$W\!=\!|\nabla V|^2$. 
It is clear that the absolute minima of $W$, located at $W\!=\!0$, 
are true saddles of the energy surface (for simplicity of notation  
we call ``saddles'' also the minima and maxima of $V$)
while the local minima of $W$ (those with $W>0$) correspond to points  
with (at least) one inflection direction, and are not saddles in mathematical sense, 
rather they are ``shoulders'' along the inflection direction.
As pointed out by
Doye and Wales \cite{doye}, the local minima of $W$, and not the  
absolute ones, are very often encountered during the minimization  
procedure. However, as it will be clear soon, the
properties of the local and absolute minima of $W$ which are actually  
important in determining the diffusive behavior are exactly the same.
For this reason we call the local minima of $W$   {\it
quasisaddles}, to emphasize the fact that they carry the same  
information as saddles, even if they are geometrically different
in nature (for a more detailed discussion see
\cite{noi_jcp,comment,rensp}).

Besides the landscape picture of the dynamic processes, the study of  
saddles has also permitted a quantitative characterization of the main  
features of the PES of liquid systems. Indeed, important
PES properties, as the mean energy elevation of saddles from underling  
minima or the Euclidean distances among saddles, can be
inferred from the analysis of saddle properties. It emerges an
high regularity of the PES, with few parameters describing the
spatial and energetic location of saddles.

In this work we apply the saddle-approach to different model liquids  
(Lennard-Jones like pair potentials), in order to better understand the  
relationship between landscape properties and slow
dynamics, and in order to evidencing the existence of general
features of the PES. The main result of this work is the existence of  
master curves both for temperature-dependent properties (saddle order  
vs. $T$) and for landscape properties (saddle energy vs. order), once  
energies and temperatures are normalized to $T_{MCT}$. This is a very  
strict relationship between dynamics and
landscape features: differences in the PES for different systems
simply define different $T_{MCT}$ values, and once scaled by these
values, one obtains exactly the same behavior. In other words, it
appears that the PESs are very similar, the only differences being
the values of few parameters describing them (like the mean
elevation barriers $\Delta E$ - mean elevation of saddles of order
one from underlying minima) that lead to different values of
dynamical quantities ($T_{MCT}$). The last point is of particular
importance: for all the systems investigated we obtain that the value  
of $T_{MCT}$ is about $1/10$ of the energy barrier $\Delta E$,
suggesting a kind of universality in the rearrangement processes  
governing the diffusion.

\section{Models}
We numerically investigated four different Lennard-Jones like model  
systems, all composed of $N\!=\!256$ particles inside a cubic box with  
periodic boundary conditions. These are:
\begin{enumerate}

\item the modified monatomic Lennard-Jones (MLJ) \cite{mlj}, at $\rho$=$1.0$
(hereafter all the quantities will be expressed in LJ reduced
units)
\begin{equation}
V_{MLJ}(r) = 4 \epsilon \left[ (\sigma/r)^{12} 
-  (\sigma/r)^6 \right] + \delta V\ ,
\end{equation} 
where $\delta V$ is a (small) many-body term that inhibits crystallization
\begin{equation}
\delta V = \alpha \Sigma_{\vec q}  \; \;
\theta ( S({\vec q}) - S_0 ) \
\left [ S({\vec q}) - S_0 \right ]^2 \ .
\label{castra}
\end{equation}
$S({\vec q})$ is the static structure factor,
the sum is made over all $\vec q$ with
$q_{max}$$-$$\Delta$$<$$|{\vec q}|$$<$$q_{max}$+$\Delta$, where
$q_{max}$=7.12$(\rho)^{1/3}$ and $\Delta$=0.34, 
and the values of the parameters are $\alpha$=0.8 and $S_0$=10.

\item the modified monatomic soft spheres (MSS), at $\rho$=$1.0$,
\begin{equation}
V_{MSS}(r) = 4 \epsilon (\sigma/r)^{12} + \delta V\ ,
\end{equation} 
where $\delta V$ is defined in Eq. \ref{castra}.

\item the binary mixture Lennard-Jones $80$-$20$ (BMLJ) \cite{bmlj},
at density $\rho$=$1.2$,
\begin{equation}
 V_{BMLJ}(r) = 4 \epsilon_{\alpha \beta} \left[ 
(\sigma_{\alpha \beta}/r)^{12} -  (\sigma_{\alpha \beta}/r)^6 \right] \ ,
\label{vbmlj}
\end{equation} 
where the values of the parameters are those of the Kob-Andersen mixture
($\sigma_{AA}\!=\!1$, $\sigma_{AB}\!=\!0.8$, $\sigma_{BB}\!=\!0.88$,
 $\epsilon_{AA}\!=\!1$, $\epsilon_{AB}\!=\!1.5$, $\epsilon_{BB}\!=\!0.5$);

\item a variant of the binary mixture Lennard-Jones (BMLJ$_2$),
at $\rho $=$1.2$, in which the values of $\sigma _{AA}$ and $\sigma _{BB}$ were exchanged.

\end{enumerate}
In the case of BMLJ and BMLJ$_{2}$, the interaction
potential
is tapered at long distances between  $r_1\!=\!2.43\sigma _{AA} \leq  
r\leq 2.56\sigma _{AA}\!=\!r_2$
with the following fifth-order smoothing function  
$\mathcal{T}(r)=1+(r_{1}-r)^{3}(6r^{2}+
(3r+r_{1})(r_{1}-5r_{2})+10r_{2}^{2}) \diagup (r_{2}-r_{1})^{5}$.
In this way the potential, the forces and their derivatives are
continous, the energy can be kept constant to better than
$1/10^{5}$ over $100$ millions of time steps. The MLJ and MSS potential  
have been simply cut and shifted at 2.5$\sigma$.

We performed standard molecular dynamics simulations at
equilibrium ($NVE$ ensemble), in a temperature range from
$T\!=\!2$ down to the lowest temperature that can be equilibrated in  
the MD run (this temperature is strongly model dependent). Along the  
equilibrium molecular dynamics trajectories at a given
temperature we analyzed the properties of i) the instantaneous  
configurations; ii) the inherent structures (minima); and iii) the
saddle configurations. About $1000$ configurations have been analyzed  
for each temperature and for each system. The inherent
structures associated to instantaneous configurations are
obtained by a conjugate-gradient minimization procedure on the
total potential energy. For saddles, a similar minimization procedure  
has been applied to the pseudo-potential $W\!=\!|\nabla
V|^{2}$. The tapering of the BMLJ and BMLJ$_{2}$ potentials allows the  
minimization procedures of both $V$ and $W$ to work correctly as  
they are not affected by small discontinuities in the derivative of $V$  
and $W$. The importance of avoiding discontinuities in order to obtain  
good $W$ minimization has been
recently underlined in Ref.~\cite{sad_5} where the  LBFGS
algorithm \cite{LBFGS} was used. However to obtain good
minimizations of $W$, even for a "small" system of $256$
particles (i.e. 768 dimensions), is a stiff problem. 
We tested different minimization algorithms
(steepest-descent, Gauss-Newton,
preconditioned conjugate gradient, Levenberg-Marquardt
\cite{numrec})
but they eventually stick in some  
points of the configuration space, where the algorithm
decrease more and more the step size, and the search becomes  
inefficient and possibly stops. Different algorithms usually stick
in different points. Sometimes the same algorithm who stuck in a given  
point can be effective in overcoming the critical situation if a larger  
step is used. Therefore, in the present work, a complex flow chart with  
various algorithms was used to obtain good
minima (the details of the numerical algorithms will be presented  
elsewhere \cite{MS1}). We want to remark that in this way the calls to  
the function $W$\ are always less than $3500$ (average
$\approx 1500$)\ and less than $1000$ (average $\approx 200$) to the  
derivative of $W$.

For all the analyzed configuration points (instantaneous, minima
and saddles) we store the energies per particle ($e$, $e_{_{IS}}$
and $e_s$ respectively), and for instantaneous and saddles we also
determine their order $n$ and $n_s$, defined as the fractional
number of negative eigenvalues of the Hessian, i.e. the absolute number  
of negative curvatures over $3N$ (for inherent structures one obviously  
has $n_{_{IS}}$=$0$).
\begin{table}
\begin{ruledtabular}
\begin{tabular}{lccrc}
Models
&$\rho$
&$T_{MCT}$
&$\Delta E$
&$\Delta E^*$ \\
\hline
MLJ & 1.0 & 0.475   & 4.43  & 9.3 \\ %9.33 \\
MSS & 1.0 & 0.210   & 2.06 & 9.8 \\ %9.84 \\
BMLJ  & 1.2 & 0.435 & 4.16 & 9.6 \\ %9.58 \\
BMLJ$_2$ & 1.2 & 0.605   & 5.93 & 9.8 \\ %9.81 \\
\end{tabular}
\end{ruledtabular}
\caption{\label{table}For the different Lennard-Jones like models
we report the investigated density $\rho$, the mode-coupling
temperature $T_{MCT}$ (estimated from the apparent power-law
vanishing of the diffusion coefficient), the mean barrier values
$\Delta E$ (mean elevation of order-one saddles from underlying
minima) and the reduced barrier height $\Delta E^* \!=\! \Delta E
/ T_{MCT}$. All the quantities are in LJ reduced units. }
\end{table}

\section{Saddles and quasisaddles}

\begin{figure}[t]
%%\epsfysize= 10 truecm
%%\begin{center}
\includegraphics[width=.5\textwidth]{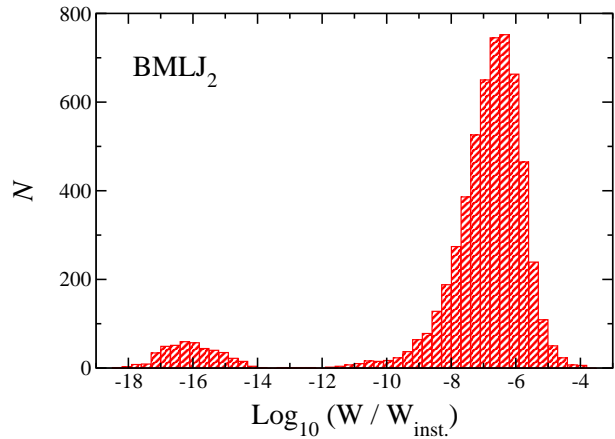}
%%\end{center}
\caption{Histogram of the ratio $W/W_{inst,}$, i.e. the value at
the minima of $W$ with respect to the value at instantaneous
configurations, at $T\!=\!2$ for BMLJ$_2$ ($6000$ configurations analyzed). 
The higher region corresponds to quasisaddles
(local minima of $W$), while the lower one to true saddles
(absolute minima).} 
\label{fig_1}
\end{figure}

First of all we focus our attention on the differences between
saddles (absolute minima of $W$) and quasisaddles (local
minima of $W$). As an example, Fig. \ref{fig_1} shows, for
the case of BMLJ$_2$ model, the histogram of the value of the
pseudopotential $W$ at the minima ($6000$ configurations analyzed at $T\!=\!2$).
The values of $W$ at the minima are
normalized to the values at the corresponding instantaneous
configurations, i.e. to the value of $W$ before starting the  
$W$-minimization procedure. We observe two very well distinct regions:  
the one with higher $W$ values corresponds to local
minima of $W$ (quasisaddles), the lower one corresponds to absolute  
minima (true saddles). The non-zero values of $W$ on the low-$W$ peak  
is due to the finite precision and/or threshold
employed in the minimization procedure. A closer inspection of the
eigenvalues of the Hessian shows that the quasisaddles are points
with only one extra zero eigenvalue \cite{comment} (besides the
three connected to the global translations), corresponding to an
inflection one-dimensional profile along the corresponding
eigenvector. The fact that in the plot the two regions are well
separated, allows to discriminate true and false saddles in a
clear way. On the contrary, no clear separation has been found
between saddles and quasisaddles from the analysis of the
eigenvalues: due to the finite precision the found eigenvalues
relative to the inflection points are different from zero of the same  
amount of the lowest frequency eigenvalues of real vibrational (or  
diffusive) modes. As it is evident from Fig. \ref{fig_1}, true-saddles are  
very few and their number are found to decrease on lowering the  
temperature (e.g. for BMLJ$_2$ in Fig. \ref{fig_1} about $5\%$ at $T\!=\!2$, 
and for BMLJ about $2\%$
at $T\!=\!2$ and less than $1\%$  at $T\!=\!0.48$).
%%

\begin{figure}[t]
%%\epsfysize= 10 truecm
%%\begin{center}
\includegraphics[width=.5\textwidth]{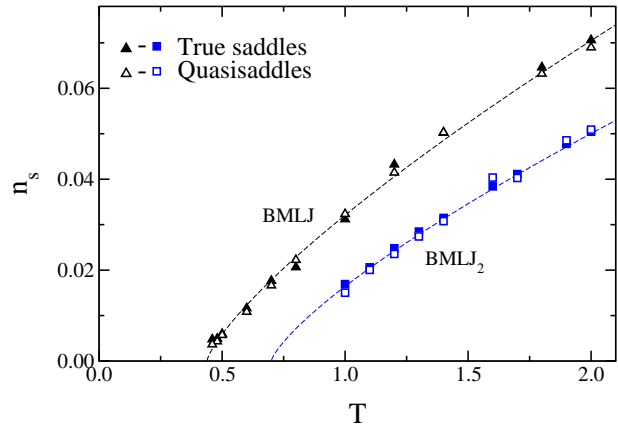}
%%\end{center}
\caption{Temperature dependence of the fractional order of true
saddles (full symbols)and quasisaddles (open symbols), for BMLJ
(triangles) and BMLJ$_2$ (squares). Dashed lines are power law
fits.} 
\label{fig_2}
\end{figure}

An interesting observation arises from the analysis of the behavior of  
the $T$-dependence of the number of negative curvatures in the "true"  
saddles and in the quasisaddles separately. 
In Fig. \ref{fig_2} the saddle order is shown as a function of the
temperature using only the true- (full symbols) and the quasi-
(open symbols) saddles, for the cases of BMLJ (triangles) and
BMLJ$_2$ (squares) models (we note that in the BMLJ$_2$ case, due
to the appearance of crystallization, the data are available only
for $T\gtrsim 1$). The coincidence between the two set of data
indicates that, as far as the temperature dependence of their
characteristics (order and energy) is concerned, quasisaddles and
true saddles share the same properties. Also other properties, as
for example the spectral features (i.e. the density of vibrational
states), of quasi-saddles and true-saddles are found
indistinguishable \cite{sadBLJ}.
This finding suggests that, no matter if saddles or quasisaddles,
the minimization of $W$ leads to points of the PES that are
relevant for a landscape-based interpretation of the slow dynamics
of the system: the order extrapolates to zero at the mode-coupling
temperature $T_{MCT}$ (see Table \ref{table} for the values of
$T_{MCT}$, estimated from diffusivity data, for the different
models), indicating that at this temperature the properties of the
landscape probed by the system manifest a kind of discontinuity
(the number of open directions, related to the saddle order, goes
to zero and the dynamical processes change their characteristics).
In other words, the minimization of $W$ seems to be a good method
to get ride of the fast degrees of freedom and to keep information
only on the slow degrees relevant for the slowing down of the
dynamics taking place in supercooled regime.

\section{General features of the PES}
We now turn our attention to the existence of common features
among the different model systems analyzed.
%%

\subsection{$T$-dependent properties}
\begin{figure}[t]
%%\epsfysize= 10 truecm
%%\begin{center}
\includegraphics[width=.5\textwidth]{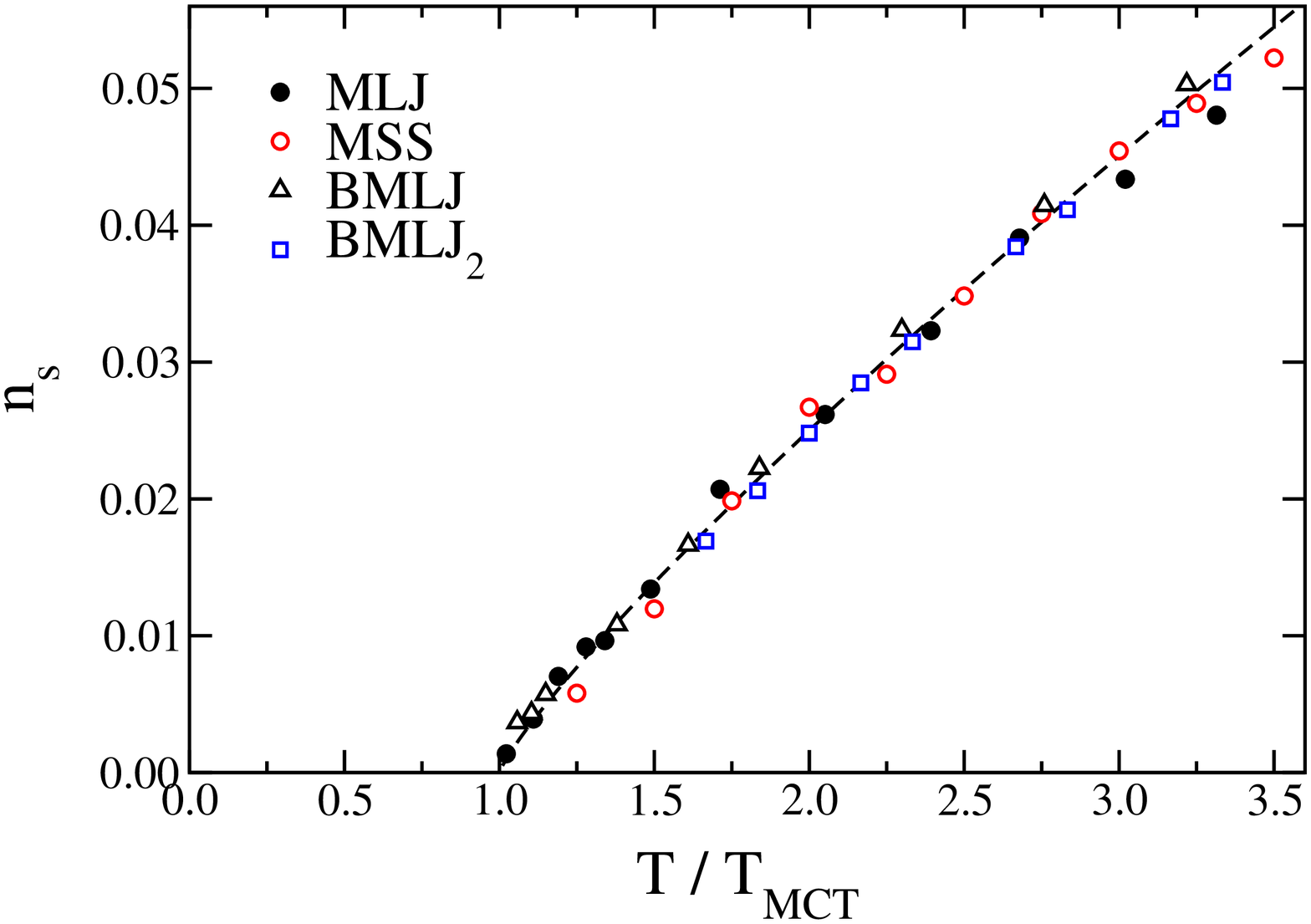}
%%\end{center}
\caption{Saddle order $n_s$ as a function of reduced temperature
$T/T_{MCT}$, for all the analyzed systems. The dashed line is a
power law with exponent $\gamma\!=\!0.85$. For MLJ and MSS
$\rho\!=\!1.0$, while for BMLJ and BMLJ$_2$ $\rho\!=\!1.2$.}
\label{fig_3}
\end{figure}

As already pointed out in Ref. \cite{noi_sad,noi_jcp}, the
(quasi-) saddle order, $n_s$, vanishes as $T$ approaches $T_{MCT}$
from above. At a first sight, it seems that the specific behavior
of $n_s(T)$ is a model-dependent property (see
Fig.~\ref{fig_2}). However, we observe that after the scaling the
temperature scale by a specific sample dependent quantity, i.~e.
by $T_{MCT}$, all the models behave similarly. In
Fig.~\ref{fig_3} the saddle order $n_s$ is reported as a function of  
reduced temperature $T/T_{MCT}$. All the curves for the different  
systems collapse into a single master curve. The latter
can be fitted by a power law
\begin{equation}
n_s = \overline{n} \left( \frac{T}{T_{MCT}} - 1 \right) ^{\gamma} \ ,
\end{equation}
with $\gamma \!=\!0.85$ and $\overline{n}\!=\!0.025$ (in the fitting  
procedure, the values of $T_{MCT}$, reported in Table\ref{table}, are  
kept fixed to the ones derived by the fit of the
power-law behavior of the diffusion coefficient). A similar master plot  
is obtained also for the relation between the saddle energy and the  
temperature. These results suggest a universal behavior
(at least for the LJ-like model systems analyzed here): at a given
reduced temperature $T^*=T/T_{MCT}$ all the systems visit saddles
with the same properties (hereafter we will indicate with ``$^*$''
the temperature and the energy scaled by $T_{MCT}$). One could
conjecture that this universality
is due to the repulsive part of
the pair potential $r^{-12}$ (common to all the systems), that
dominates over the attractive one at the studied densities.
However, the facts that the curves superimpose each other quite well in  
the whole temperature range and that non-LJ systems 
(as, for example, the Morse potential - see the next section)
show a similar behavior,
seems to indicate that the
observed universality is not trivially related to the repulsive  
part of the interaction potential. Finally, we want to
remark that the small value of $\overline{n}$ indicates that even
at temperature twice that of the MCT critical point, the system is
visiting saddles of low order ($n_s \simeq 0.025$) , so
indicating that at $T\!=\!2 T_{MCT}$
the closest saddle,  according to the partitioning
defined by the minimization of $W$, is far below the top of the  
landscape.\\
%%

\subsection{Energy barriers and $T_{MCT}$}
\begin{figure}[t]
%%\epsfysize= 10 truecm
%%\begin{center}
\includegraphics[width=.5\textwidth]{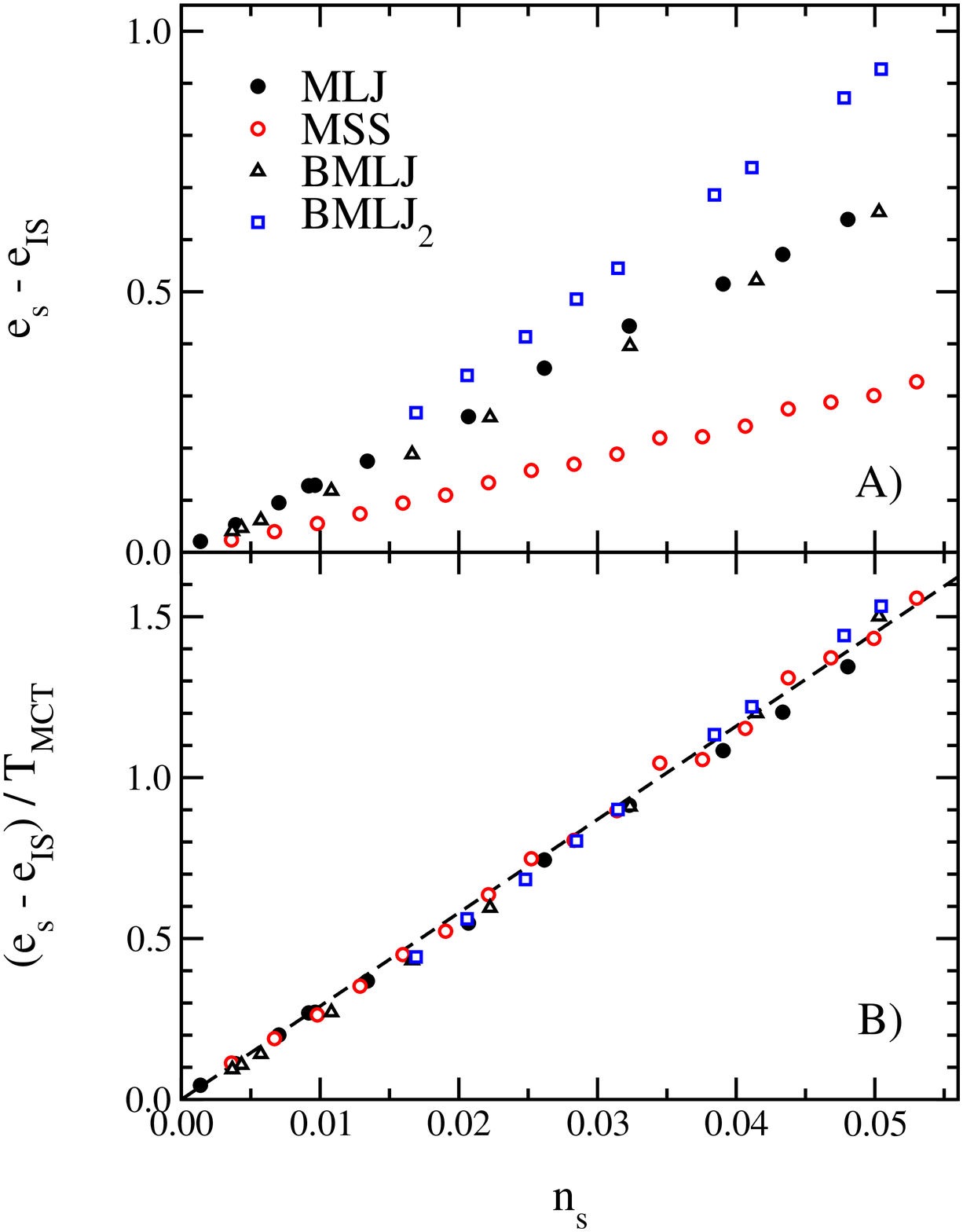}
%%\end{center}
\caption{ A) Energy elevation of saddles from underling minima
$e_s - e_{_{IS}}$ against saddle order $n_s$; B) Energy elevation
rescaled by mode-coupling temperature $T_{MCT}$ against saddle
order $n_s$. Dashed straight line is a guide to the eyes.}
\label{fig_4}
\end{figure}

The existence of common and general features of the PES emerges in
a clear way from the comparative analysis of the energy and of the
order of the saddles. In Fig.~\ref{fig_4} part A the energy
elevation $\Delta e_s$=$e_s - e_{_{IS}}$ of the saddles from the
underling minima is plotted as a function of the saddle order
$n_s$  for the different investigated models. As already observed
\cite{noi_sad}, there exists a proportionality  between these two
quantities, indicating a simple organization of the PES: saddles
are equally spaced in energy over the minima. The slopes of the
different straight lines in Fig.~\ref{fig_4} determine the
elementary energy elevation $\Delta E$ of saddles of order $n$
from saddles of order $n-1$:
\begin{equation}
\Delta E = \frac{1}{3} \frac{d (e_s - e_{_{IS}})}{d n_s} \ ,
\label{delta}
\end{equation}
where the factor $3$ is due to the fact that energies are per
particles ($N$) and the fractional order per degrees of freedom
($3N$).The values of $\Delta E$ obtained for the various systems
are reported in Table \ref{table}. A possible explanation of the
linear relationship observed in Eq.~\ref{delta} is that there
exist in the system several spatially uncorrelated rearranging
regions, each experiencing a mean barrier energy $\Delta E$. In
other words, if the system as a whole lies on a saddle of order
$m$, this is due to the fact that there are $m$ uncorrelated
subsystems each one visiting a saddle of order 1. The analysis of
the specific atomic motion associated to these saddles, needed to
assess or disprove the validity of this hypothesis, is beyond the
aim of the present work.
%%

A very interesting and surprising result is obtained by scaling
the energy values reported in Fig.~\ref{fig_4} part A to the
mode-coupling temperature $T_{MCT}$ ($K_B=1$), obtaining again a
single master curve (see Fig.~\ref{fig_4} part B). The landscapes
of different systems seem to share common features, with only one
parameter describing the organization of saddles, i.~e. the mean
elevation $\Delta E$, that becomes an universal parameter ($\Delta
E^* \!=\! \Delta E / T_{MCT}  \simeq 9 \div 10$) once normalized
to the mode-coupling temperature (see the last column of Table
\ref{table}). In other words, all the models have the common
property that the elementary barrier height is about $10$ times
the critical temperature $T_{MCT}$:
\begin{equation}
\Delta E \simeq 10\ T_{MCT} \ .
\label{univ}
\end{equation}
This relation have been numerically proved for the four potential
models investigated here. The same relation also holds for another
LJ-like model, the binary mixture soft-sphere model (BMSS)
investigated in Ref.~\cite{grig} (at $\rho=1$). This observation
gives further support to the universality of Eq.~\ref{univ}. If
this is a particular characteristic of Lennard-Jones like models
or a general feature of a more wide class of simple liquids is a
open and interesting question which remains to be answered.

We can try to give a first answer to this question analyzing the  
available data in the literature for other systems. To our knowledge,  
besides the Lennard-Jones like systems, a saddle-based analysis has  
been performed only for the Morse potential \cite{sad_4}.
The Morse potential, used in Ref.~\cite{sad_4}, is defined as
$V_\alpha(r)= \epsilon [ 1 - \exp(\alpha(1-r/r_e))]^2 -
\epsilon$, where $\epsilon$ is the well depth, $r_e$ is the  
interparticle distance and the parameter $\alpha$ is
inversely correlated to the range of the potential.
Differently from soft-spheres and LJ, the Morse potential
is finite as $r \rightarrow 0$.
Unfortunately, equilibrium simulation based on the Morse potential are  
difficult, since the  undercooled system  crystallizes easily.  
Therefore simulations reported in Ref.~\cite{sad_4} have been performed  
only well above $T_{MCT}$. It was found that the
larger the  value of $\alpha$ is , the further the distance between  
the temperature of the lowest non-crystalline simulation and $T_{MCT}$  
is. In
this study, a linear dependence between $n_s$ and $e_s$
have been observed and the values of $de_s/dn_s$ normalized to
$T_{MCT}$ are in agreement with Eq. \ref{univ} for the three smaller  
$\alpha$ values $\alpha=4,5$ and $6$ ($\Delta E^*$ are in
the range $9.3 \div 10.5$ \cite{nota_morse}). For the two highest
$\alpha$ values, $\alpha=9$ and $12$, the reported values for
$\Delta E^*$ are quite different ($4.6$ and $3.5$ respectively).  
Further studies, for example focussing on binary mixture systems,  are  
requested to
find out if  such discrepancy is due to
an approximate determination of $T_{MCT}$ for $\alpha=9$ and
$12$ (which was obtained by extrapolating $n_s(T)$ from a temperature 
region where $n_s$ is far away from zero, $n_s(T) \gtrsim 0.2$). 
Uncertanties in the estimates of $T_{MCT}$  
at large $\alpha$ are also consistent with the unexpected non-monotonic  
dependence of $T_{MCT}$ with $\alpha$ reported in Ref.~\cite{sad_4}. We  
conclude that, for all $\alpha$ values for which the reliability
of the data is unquestionable, the
Morse potential landscape  shares the same characteristic of those of  
the LJ-like potentials.

In all the other model systems studied in the literature we do not have  
a direct information on the saddle energy elevation. 
However, the existence of a well defined barrier energy scale  
$\Delta E$ in
the PES is expected to control the activation processes at low  
temperature, giving rise to an Arrhenius behavior of the transport
properties at temperatures below $T_{MCT}$. The existence of
Arrhenius law in LJ-like systems - that are basically ``fragile'',
in the Angell classification scheme \cite{land_angell} - would be,
per se, surprising (however, the degree of fragility of LJ systems
is a matter of debate \cite{angell_pisa}).
\begin{figure}[t]
%%\epsfysize= 10 truecm
%%\begin{center}
\includegraphics[width=.5\textwidth]{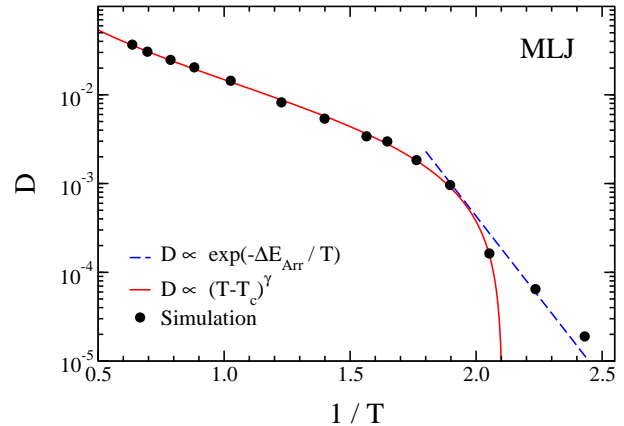}
%%\end{center}
\caption{Diffusivity $D$ as a function of inverse temperature
$1/T$ for the MLJ model. Straight line represents the mode
coupling like power law fit. Dashed line is 
the Arrhenius law with energy barrier $\Delta E_{Arr}
\simeq 1.9 \Delta E = 8.4$ ($\Delta E$ is the energy barrier 
from saddles - see Table \ref{table}), following the corresponding relation obtained 
for the BMLJ case (Ref. \cite{sastry_pisa}).}
\label{fig_5}
\end{figure}%%

The simulations below $T_{MCT}$ are very
difficult to perform, due to the extremely long relaxation times
in this regime and a direct inspection of the expected Arrhenius
behavior is not easy to pursuit. Only very recently such a kind of
analysis has been performed for the BMLJ model at
$\rho=1.2$ \cite{sastry_pisa}. In that work an Arrhenius behavior
was actually found in the temperature dependence of the diffusion
coefficient below $T_{MCT}$ : $D \propto \exp(-\Delta E_{Arr} /T)$
(we use the symbol $\Delta E_{Arr}$ for the activation energy in
the Arrhenius law of the diffusivity, to distinguish it from the
energy barrier $\Delta E$ determined from the saddles analysis of
the PES), with a value of $\Delta E_{Arr} \simeq 8.1$. The
observed Arrhenius behavior is somewhat surprising in this
``fragile'' liquid models, and seems to indicate that close to
$T_{MCT}$ activated processes start to be relevant and dominate
the dynamics. However, the value of the activation energy $\Delta
E_{Arr}$ found in Ref.~\cite{sastry_pisa} is not equal to the
elementary barrier energy $\Delta E$ estimated from the saddles
analysis (see Table \ref{table}), but it is about twice that
value: $\Delta E_{Arr}/\Delta E \simeq 1.9$. Re-analyzing our data
for the MLJ model (for which we have few thermodynamic points
equilibrated close to but below $T_{MCT}$), we find that the above
reported ratio is compatible with MLJ data (see Fig.~\ref{fig_5}),
even if statistic is poorer than that of BMLJ case and the
equilibrium condition is not fully satisfied by the lowest two
temperature points. If such observation has general validity, then
Arrhenius behavior should be observed below $T_{MCT}$, with an
activation energy value about $2$ times the value of the elementary  
saddle energy barrier [so obtaining a value of reduced barrier
energy (normalized to $T_{MCT}$)  $\Delta E_{Arr}^* \simeq 18
\div 20$]. The origin of this factor two needs to be further clarified.  
To this aim, it is important to underline that the
"effective" energy barriers for activated processes as seen by the
dynamics (i.~e. those entering in the Arrhenius law for the
diffusivity) can be higher than the minimum-to-saddle energy
difference (as measured directly by analyzing the PES). This can
be due to the fact that the true diffusive path in the landscape
\cite{demic} could pass higher in energy with respect to the
saddle point, in order, for example, to minimize the
minimum-to-minimum path length (i.~e. for entropic reasons). In
this respect, it is worth to mention that a non-coincidence between
the relaxation times determined either through MD simulations or
through the direct inspection of the PES has been observed in the
simulation of a model protein during the folding process
\cite{torcini}. In particular, the results in Ref.~\cite{torcini}
indicates that the effective saddle height is larger than the
actual one.
\begin{table}
\begin{ruledtabular}
\begin{tabular}{lcc}
Models
&$\Delta E^*$
&$\Delta E_{Arr}^*$ \\
\hline
MLJ & 9.3 & 17.7 \\ %9.33 \\
MSS & 9.8 & ... \\ %9.84 \\
BMLJ  & 9.6 & 18.6\footnote{Ref. \cite{sastry_pisa}} \\ %9.58 \\
BMLJ$_2$ & 9.8 & ... \\ %9.81 \\
BMSS\footnote{Ref. \cite{grig}} &  9.1 & ... \\
Morse\footnote{Ref. \cite{sad_3} (obtained for
$\alpha\!=\!4,5,6$)}  &  $9.3\div10.5$ & ... \\ Silica
(BKS)\footnote{Ref. \cite{horbac}} & ... & 16$\div$18 \\
OTP\footnote{Ref. \cite{mossa}} & ... & $20\div28$ \\ Water
(SPC/E)\footnote{Ref. \cite{starr}} & ... & 40 \\
\end{tabular}
\end{ruledtabular}
\caption{\label{table2} Reduced energy barrier heights estimated
from saddles ($\Delta E^*\!=\!\Delta E / T_{MCT}$) and from
low-temperature Arrhenius law of diffusivity ($\Delta
E_{Arr}^*\!=\!\Delta E_{Arr} / T_{MCT}$) for different model
systems. The data of MLJ, MSS, BMLJ and BMLJ$_2$ are from this
work (except the $\Delta E_{Arr}^*$ for BMLJ, that is from Ref.
\cite{sastry_pisa}). }
\end{table}

Having in mind that $\Delta E_{Arr}^* \! \simeq \! 2 \Delta E^*$
and that $\Delta E^*\! \simeq\!10$ (i.e. $\Delta E_{Arr}^*
\!\simeq\! 20$), we can try to analyze what is observed for other
model potentials existing in the literature where the $D(T)$ has
been determined. We found three different models for which a low
temperature analysis of $D(T)$ has been performed {\it via}
molecular dynamics: {\it i)} The BKS-silica model \cite{horbac},
for which the values of $\Delta E_{Arr}^*$ are $16.2$ and $18.0$,
for the self diffusion of $O$ and $Si$ respectively; 
{\it ii)} the Lewis and Wahnstr\"om ortho-terphenyl  
model \cite{mossa}, for
which the temperature dependence of the molecular center of mass
diffusion coefficient at five different densities give values of
reduced barrier energy $\Delta E_{Arr}^* \simeq 20 \div 28$
(except the lowest density that gives a value of about $10$); {\it
iii)} The SPC/E-water model \cite{starr}, for which one finds
$\Delta E_{Arr}^* \simeq 40$. Table \ref{table2} summarizes the
known results on energy barrier heights estimated from saddles and
from Arrhenius low-temperature dependence of diffusivity. The
values for MLJ, MSS, BMLJ and BMLJ$_2$ are from the present work,
except the $\Delta E^*_{Arr}$ for BMLJ that is obtained from Ref.
\cite{sastry_pisa}. In future works we will try to determine the
saddle-barriers $\Delta E^*$ for non-LJ systems (the last three
systems in the Table), in order to have a better understanding of
the diversity of the different landscapes. In conclusion, besides
the case of water, the other systems seem to be in agreement with
the findings of this work (the values of the reduced barrier
energies are of the same order), evidencing a quite general
universality of the observed relations. A deeper understanding of
the differences among various model liquids deserves further
investigations.

\section{Conclusions}
In conclusion, despite complex and disordered in nature, the
simple liquid PES seems to exhibit few general and regular
features, useful both to bring important insight for the
understanding of the relevant diffusion processes taking place in
supercooled liquids and to construct simplified PES models. The
main findings of the present work can be summarized as:
\begin{itemize}
\item the coincidence between the temperature dependence of the
quasisaddles and of the true saddles properties;\\
\item the existence of master curves for saddle properties, once  
energies
and temperatures are rescaled by the mode coupling critical
temperature $T_{MCT}$;\\
\item the existence of a universal relationship between the  
mode-coupling
temperature and the mean energy barrier height $\Delta E \simeq
10\ T_{MCT}$, that seems to extend beyond the class of the
Lennard-Jones like models analyzed here.\\
\end{itemize}

Finally, we would like to point out that it already exists in the  
literature an hint on the existence of a linear relationship
between  $\Delta E_{Arr}$ and the mode coupling critical
temperature. Indeed, in a large class of glassy system one
experimentally observes a linear relationship between the glass
transition temperature $T_g$ and the infinite-frequency shear
modulus $G_{\infty}$ \cite{nemilov}: $T_g \propto G_{\infty}$. If
we use the findings of our work ($\Delta E_{Arr}^* \simeq 10$,
i.e. $\Delta E_{Arr} \simeq 10\ T_{MCT}$) and we allow ourselves to
confuse $T_g$
with $T_{MCT}$, the following relation emerges: $\Delta E_{Arr}
\propto G_{\infty}$. This relation is the prescription of the
``shoving model'' 
introduced thirty years ago by Nemilov \cite{nemil_68} and
recently put in a more rigorous form
by Dyre {\it et al.}  \cite{dyre}. 
The validity of the
proportionality between $\Delta E_{Arr}$ and $G_{\infty}$ has been
proved for different glasses, and, together with the
linear relationship between $T_g$ and $G_{\infty}$, give further
support to the finding of the present work, i.~e. the apparent  
universality of the ratio $\Delta E_{Arr}/T_{MCT}$.
\\

\begin{acknowledgments}
We acknowledge support from INFM, PRA GenFDT,  MURST COFIN2002 and
FIRB. 
G.R. thanks J.C.~Dyre for useful discussions.
\end{acknowledgments}


\begin{thebibliography}{99}
\bibitem{nota_morse}
The data are obtained from Table I in Ref. \cite{sad_4}, dividing
the values in the second column by a factor $3$, for the same
reason as reported in the text - see Eq. \ref{delta}.
\end{thebibliography}
\end{document}